\documentclass{article}
\usepackage{spconf,amsmath,graphicx}
\usepackage[hidelinks]{hyperref}
\usepackage{siunitx}
\usepackage{amsfonts}
\usepackage{multirow}
\usepackage{makecell}
\usepackage{booktabs}

\DeclareSIUnit{\pp}{pp}

\title{TOWARDS MICROPHONE-INVARIANT SPEECH REPRESENTATIONS FOR TRANSFORMER-BASED VOICE DISORDER DETECTION}

\name{Iva Ewert \qquad Felix Putze \qquad Marvin Borsdorf \qquad Tanja Schultz}
\address{Cognitive Systems Lab, University of Bremen, Germany}

\begin{document}
\ninept
\maketitle
\begin{abstract}
Automatic voice disorder detection from speech offers strong potential for early diagnosis and continuous disease monitoring.
While models based on \mbox{pre-trained} transformers have improved detection performance, their robustness to microphone variation remains underexplored, particularly for \mbox{consumer-grade} devices.
To the best of our knowledge, we present the first study of microphone mismatch in \mbox{transformer-based} binary voice disorder detection across \num{20} smartphone microphones. 
While most mismatched microphone conditions cause moderate performance degradation, the \mbox{worst-case} condition reduces unweighted average recall by up to $\num{12.53}$ percentage points.
To address this, we introduce a Barlow \mbox{Twins-inspired} module to learn \mbox{microphone-invariant} representations. 
Our approach reduces the performance gap between matched and mismatched conditions by up to \num{62}\%, 
while maintaining comparable \mbox{matched-condition} performance. 
Qualitative analysis of the embedding space demonstrates more compact, distinct speaker clusters across microphones. 
Importantly, results indicate generalisation to unseen microphones across manufacturers, smartphone generations, and release periods, without requiring any \mbox{inference-time} adaptation. 
\end{abstract}
\section{Introduction}
\label{sec:intro}

Speech is a promising biomarker for the early detection and continuous, non-invasive monitoring of
several diseases, as the speech signal is influenced through diverse physiological mechanisms~\cite{botelho2024speech}. 
Leveraging this potential, numerous studies have demonstrated the utility of speech recordings for the automatic detection of diverse health issues, including Alzheimer's dementia~\cite{weiner}, depression~\cite{menne2024voice}, as well as voice disorders~\cite{vidal2024automatic}.
This work focuses on detecting the presence of voice disorders, i.e. conditions that disrupt the normal production or function of the human voice.
Voice disorders are prevalent among occupational groups that frequently use their voice~\cite{vilkman2000voice}
and can substantially limit daily activities and social participation of affected individuals~\cite{ma_voice_2001}.

Machine learning approaches have been developed to distinguish pathological voices, i.e., voices affected by a voice disorder, from healthy voices~\cite{verde2018voice,Koudounas2024-ch,medjani2025automatic}, or to detect specific types of voice disorders~\cite{wang2022continuous, Peng_Xu_Liu_Wang_He_2023} based on speech recordings.
While traditional methods use \mbox{hand-crafted} features like Mel Frequency Cepstral Coefficients,
latest studies directly apply deep neural networks (DNNs)  
to raw audio.
Recent works~\cite{Koudounas2024-ch, medjani2025automatic} have explored the use of 
transformer-based models like Wav2Vec~\cite{baevski2020wav2vec} or HuBERT~\cite{hsu2021hubert}, 
pre-trained on huge amounts of unlabeled speech data to yield rich contextualized speech representations by \mbox{self-supervised} learning tasks.
\mbox{Pre-trained} models have advanced the field of 
voice disorder detection and outperform \mbox{hand-crafted} feature based approaches~\cite{ribas2023automatic}.

However, \mbox{real-world} deployment
remains challenging since recording conditions often differ substantially from those in the controlled clinical training data.
For example, when applied
to smartphone data, models must generalise well across 
diverse, potentially low-cost,
microphones and remain robust to changes in recording technology.
In other speech domains,
studies have demonstrated that microphone mismatch can adversely affect the performance of \mbox{transformer-based} speech models~\cite{wang2025channel}.
In voice disorder detection, Schaeffler et al.~\cite{schaeffler2019reliability} reported systematic differences
in acoustic voice parameters when comparing simultaneous recordings obtained with a studio microphone and
smartphone microphones.
Moreover, recent work has reported performance degradation of \mbox{DNN-based} voice disorder detection models, with reductions of up to $\num{9.5}$ percentage points
in the unweighted average recall~(UAR)~\cite{wang2026domain}.
However, to the best of our knowledge, the impact of microphone mismatch on transformer-based voice disorder detection models has not yet been investigated. 
Furthermore, existing work focuses on professional microphones~\cite{wang2026domain}, leaving \mbox{consumer-grade} microphones, as those embedded in modern smartphones, largely unexplored.
Beyond characterizing the impact of microphone mismatch on voice disorder detection models, developing effective strategies to improve robustness remains a persistent challenge.
Prior works focused on \mbox{domain-adversarial} approaches~\cite{wang2026domain,Hsu2018RobustnessAT}, but despite their demonstrated successes, they are often hampered by
training instability and the risk of degenerated representations. 
This work contributes to closing the identified gaps by:
\begin{enumerate}
  \item  Evaluating the performance degradation of transformer-based voice disorder detection models under microphone mismatch, with a focus on smartphone devices.
  \item Proposing the integration of a Barlow Twins~\cite{zbontar2021barlow} inspired module to promote microphone invariance and prevent representational collapse via redundancy reduction.
  \item Reducing the performance gap between matched and mismatched microphone conditions significantly for most considered smartphone microphones, while maintaining comparable \mbox{matched-condition} performance, and without requiring any adaptation at inference time. 
  \item Showing refined embedding space organization with more compact, distinct speaker clusters across microphones. 
\end{enumerate}

\section{Dataset and microphone simulation}

To assess the robustness of voice disorder detection across different recording devices, we apply realistic microphone characteristics to real pathological speech.
Following the strategy of Wang et al.~\cite{wang2026domain}, we simulate microphones by convolving pathological speech with device impulse response functions (DIRs).
Specifically, we use DIRs from the Poliphone dataset~\cite{salvi2025poliphone} captured for diverse smartphones. 
 
\subsection{Device impulse response functions}
\label{subsec:dirs}

The Poliphone~\cite{salvi2025poliphone} dataset contains DIRs for \num{20} smartphone models from ten manufacturers, launched between 2018 and 2022.
Detailed information on the
devices is provided in the original paper~\cite{salvi2025poliphone}.
DIRs were extracted from chirp recordings in an anechoic chamber under consistent conditions, minimising environmental influence. 
Convolving a DIR with an audio signal provides it with the recording characteristics of that specific microphone.

\subsection{Pathological voice dataset}

We use the Saarbrücken Voice Dataset~\cite{woldert_jokisz2007saarbruecken} for binary voice disorder detection, assigning all pathological voices (\num{69}\%) to one class and all healthy voices to the other. 
The dataset contains speech recorded with a headset condenser microphone under controlled conditions~\cite{putzer2008instrumental}
and we use all sentence recordings.
The data is split at the subject level, with \num{20}\% of speakers (\num{323}) held out for testing and the remaining \num{80}\% split \num{80}/\num{20} into training and validation.
All audio is downsampled to \qty{16}{\kilo\hertz},
segmented into \mbox{non-overlapping} \mbox{three-second} chunks, and \mbox{zero-padded} if shorter. Samples shorter than one second are discarded, and all samples are normalized.

\section{Modeling framework}
\label{sec:modeling_framework}
\begin{figure}
  \centering
  \includegraphics[width=\columnwidth]{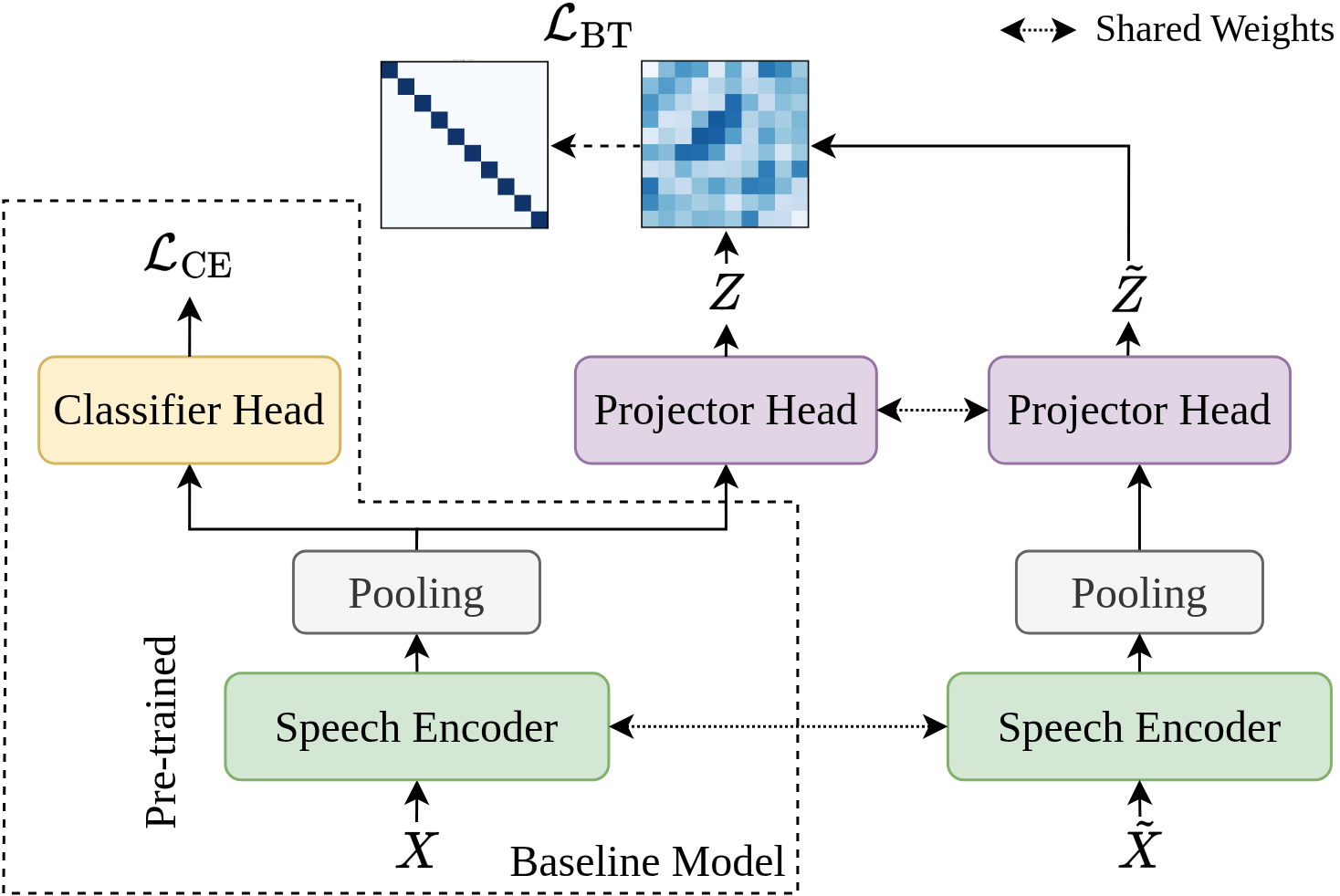}
  \caption{The proposed model consists of a \mbox{pre-trained} speech encoder, a pooling layer, a classifier head, and a projector head. 
  In addition to the standard \mbox{cross-entropy} loss $\mathcal{L}_{\text{CE}}$ (applied to the classifier output), we apply a Barlow Twins (BT) loss ($\mathcal{L}_{\text{BT}}$) to the pair of projector outputs denoted as $Z, \tilde{Z} \in \mathbb{R}^{B \times D}$. The pair is computed by passing a batch of speech samples $X$ and its microphone transformed version $\tilde{X}$ through the same model. 
  The baseline model (dashed) does not include the projector head and uses only the classification loss. 
  At inference time, the BT model requires only the untransformed test samples $X$ and the classification branch to obtain the prediction.}
  \label{fig_modeling_framework}
\end{figure}

The baseline, indicated by a dashed line in Figure~\ref{fig_modeling_framework}, 
comprises a \mbox{pre-trained} transformer-based speech encoder, followed by mean pooling, and a simple classifier head.
The classifier head consists of two dense layers: a \mbox{\num{256}-unit} hidden layer with ReLU activation, followed by an output layer. 

\subsection{Pre-trained speech encoder}

For the speech encoder, we consider two pre-trained models: Wav2Vec~\cite{baevski2020wav2vec} (``wav2vec2-base''\footnote{https://huggingface.co/facebook/wav2vec2-base}) as well as HuBERT~\cite{hsu2021hubert} \linebreak (``hubert-base-ls960''\footnote{https://huggingface.co/facebook/hubert-base-ls960}).
Both architectures have been successfully applied to voice disorder detection \cite{medjani2025automatic,Koudounas2024-ch, B2026103464}, demonstrating their ability to capture pathological speech characteristics.
Both consist of a convolutional feature encoder followed by a transformer~\cite{vaswani2017attention} for contextualized speech representations and were pre-trained on \mbox{\num{960} hours} of speech from the \mbox{LibriSpeech} corpus~\cite{panayotov2015librispeech}.
The key difference is their \mbox{self-supervised} learning \mbox{pre-training} objective: Wav2Vec learns to predict masked latent representations, whereas HuBERT learns to predict masked pseudo-labels. 
We include both to assess whether our findings generalise across \mbox{pre-training} strategies.

\subsection{Barlow Twins loss}
\label{sec:bt_loss}

To improve microphone robustness, we incorporate a Barlow Twins (BT)~\cite{zbontar2021barlow} style alignment module.
This module applies a projector head (see Figure~\ref{fig_modeling_framework}) to the pooled embeddings, projecting them into a space where the BT objective is enforced.
The projector head consists of a \mbox{\num{512}}-unit hidden layer followed by batch normalization and ReLU activation, and a final layer projecting to a \mbox{\num{256}-dimensional} representation.
The BT loss~\cite{zbontar2021barlow} is defined as 
\begin{equation*}
  \mathcal{L}_{\text{BT}} = \sum_{i} (1 - C_{ii})^{2}+ \lambda \sum_{i}\sum_{j \neq i} {C_{ij}}^{\,2},
\end{equation*}
where  
$C \in \mathbb{R}^{D \times D}$ is the \mbox{cross-correlation} matrix between the projector embeddings $Z, \tilde{Z} \in \mathbb{R}^{B \times D}$ of the original and transformed input batches ($X, \tilde{X}$), respectively,
with $B$ being the batch dimension and $D$ the embedding dimension.
The first term of the loss drives the \mbox{cross-correlation} matrix towards identity, promoting invariance of the embeddings under the applied transformation, which in this work corresponds to microphone simulation.
The second term penalizes correlation between different embedding dimensions, thereby promoting redundancy reduction.
This term is weighted by $\lambda = 0.005$,
as suggested by Zbontar et al.~\cite{zbontar2021barlow}.
The total loss of the BT 
model is given by 
\begin{equation*}
  \mathcal{L} = \mathcal{L}_{\text{CE}} + \alpha \mathcal{L}_{\text{BT}},
  \label{eq:total_loss}
\end{equation*}
where $\mathcal{L}_{\text{CE}}$ is the standard cross-entropy loss.
The loss weight $\alpha$ is chosen as \num{0.05} to 
balance the initial magnitudes of the two loss components.
The baseline model is trained using only $\mathcal{L}_{\text{CE}}$.

In a separate experiment, the baseline model is trained with data augmentation to compare its performance against the BT approach. 
Here, each training sample is uniformly assigned to either no transformation
or to one of the microphone transformations from the given device subset.

\section{Experimental setup}
\label{sec:exp_setup}

This section outlines the training specifications, the design of device subsets, and the evaluation pipeline.

\subsection{Model training}

All components, including the classifier (and projector) head as well as the pre-trained speech encoders, are trained jointly using the objectives in Section~\ref{sec:bt_loss}. 
We use Adam~\cite{kingma2014adam} with batch size \num{32}, early stopping, and a maximum of \num{50} training epochs.
Learning rate and weight decay are selected via Optuna's Bayesian optimization~\cite{akiba2019optuna} (\num{20} trials, \num{10} epochs each) based on validation performance.
Starting from \mbox{pre-trained} encoder weights, we apply a linear learning rate warm-up over the first \num{5} epochs, starting at \num{10}\% of the final rate.
The best model is selected by UAR on the validation set, and a \num{10}\% dropout is applied to the classifier head.
Experiments use PyTorch on a single NVIDIA RTX 6000 Ada Generation GPU, with each experiment repeated \num{20} times using different random seeds,
and the results are reported as aggregated values.

\subsection{Device subsets}
\label{sec:device_subsets}

To evaluate the approach's generalisation to unseen microphones and its sensitivity to the composition of the training device subset, we introduce five device splits, defining training (seen) and \mbox{held-out} (unseen) devices. 
The first split assigns one randomly chosen device per manufacturer to training (subset OnePerMfr), while excluding one manufacturer (Motorola) entirely, thereby testing within- and \mbox{cross-manufacturer} generalisation.
Split two assigns the \mbox{first-generation} smartphone for each manufacturer (if multiple generations exist) to training (subset FirstGen) and holds out the rest, 
evaluating generalisation across generations. 
The third split assigns all devices launched before 2022 to training (subset Pre2022) and holds out later devices, assessing temporal generalisation.
Split four puts all devices to training except OnePlus Nord 2 (subset \mbox{AllButOne}), which is identified as an outlier based on its DIR similarity to the other devices.
Finally, the fifth split assigns only devices from one manufacturer to training (subset OneMfrOnly), assessing cross-manufacturer generalisation under a restricted training regime.
We select Xiaomi because its DIR cluster is the most distant from those of the other manufacturers (excluding the OnePlus outlier).

\subsection{Evaluation}

At test time, predictions for speakers with multiple chunks are aggregated by averaging the logits and we use UAR for evaluation, suitable for imbalanced data.
We define the performance gap as the difference in model performance between the untransformed (original) and transformed test conditions. 
This measures the extent to which a model's performance degrades under the considered microphone transformed condition.
The prediction flip rate 
quantifies the fraction of labels that change between conditions.
To evaluate the robustness of models in comparison to the baseline model, we compute both absolute and relative reductions in performance gaps, measuring how much of the baseline gap is recovered.
Differences in performance gaps between two models are tested for significance using an unpaired Wilcoxon \mbox{rank-sum} test~\cite{wilcoxon1945individual} ($p < 0.05$), with Benjamini-Hochberg correction~\cite{benjamini1995controlling} for multiple comparisons.
For the embedding space analysis, we apply the t-distributed stochastic neighbour embedding (\mbox{t-SNE}) algorithm~\cite{van2008visualizing}  (with a maximum of \num{2000} iterations and early exaggeration of four) to project the \mbox{post-pooling} embeddings to 2D. For the visualisations, ten speakers are randomly selected from the test set using a fixed seed across models. Resulting structures are verified to be consistent across multiple t-SNE perplexities and the results shown are obtained using a perplexity of \num{30}.

\section{Results}
\label{sec:results}

The Wav2Vec and \mbox{HuBERT-based} baselines achieve \num{84.01}\% and \num{84.22}\% UAR, respectively, on the untransformed test condition.
These results are on par with 
the UAR reported by the closest prior approach (\num{84.01}/\num{85.04}\%)~\cite{ribas2023automatic}, validating the baseline for our robustness analysis.

\subsection{Robustness analysis}

The baselines show an average performance gap (UAR) of \linebreak \mbox{\num{2.68} percentage points~(\si{\pp}) (Wav2Vec)} and \mbox{\qty{2.53}{\pp} (HuBERT)} between matched and mismatched microphone conditions, accompanied by a prediction flip rate of \num{10}\% (\num{7}\% HuBERT) and a bias towards the majority (pathological) class.
While most conditions cause moderate gaps, the microphone condition corresponding to the Moto G9 device, leads to a gap of up to \SI{12.53}{pp}, likely due to strong attenuation in the \qtyrange[range-units=single]{0.5}{2}{\kilo\hertz} band 
for this microphone.
Compared to Wang et al.~\cite{wang2026domain}, where a DNN based model trained from scratch was tested on mismatched studio microphones, our gaps are smaller for most mismatched conditions.

\begin{table}[]
  \label{performance_tab}
\centering
\caption{Performance differences ($\Delta$ UAR  $\uparrow$) of the proposed Barlow Twins (BT) 
models compared to their respective baselines. The device set used for the development of each BT model is indicated, with the proportion of seen devices out of the \num{20} devices in total given in parentheses (see Section~\ref{sec:device_subsets} for details on the device splits). 
Values denote absolute percentage points (\si{\pp}) changes and are reported for the untransformed (\textit{Base}) test condition, the microphone condition with the largest baseline drop (\textit{Worst}), and averaged across all microphone conditions (\textit{All}).
For the latter condition, the reduction in standard deviation across microphone conditions relative to the respective baseline is given in parentheses.
Results for the \textit{Worst} condition are shown underlined if the respective microphone condition is seen during training. 
The \mbox{Wav2Vec/HuBERT-based} baselines achieve $\num{84.01}$/$\num{84.22}$\% on the untransformed (\textit{Base}) condition,
$\num{71.48}$/$\num{72.55}$\% on the \textit{Worst} condition and $\num{81.33}$/$\num{81.69}$\% across all conditions (\textit{All}).}
\label{tab:mic_robustness}

{\fontsize{9pt}{10pt}\selectfont
\setlength{\tabcolsep}{4.5pt}
\begin{tabular}{@{}r l||rrl}
&
&
\multicolumn{3}{l}{\textbf{Microphone condition}}
\\
&
\textbf{Train device subset}
&
\textit{Base}
&
\textit{Worst}
&
\textit{All}
\\
\midrule
\multirow{5}{*}{\rotatebox{90}{\textbf{Wav2Vec}}}
& OnePerMfr (10/20) & $+\num{0.16}$ & $+\num{7.07}$  & $+\num{1.81}$ ($-\num{55}$\% std.)\\ 
& FirstGen (5/20) & $-\num{0.56}$ & \underline{$+\num{8.52}$}  & $+\num{1.18}$ ($-\num{71}$\% std.) \\ 
& Pre2022 (17/20) & $-\num{0.08}$ & \underline{$+\num{8.26}$}  &  $+\num{1.56}$ ($-\num{65}$\% std.) \\ 
& AllButOne (19/20) & $-\num{0.40}$  & \underline{$+\num{7.88}$} & $+\num{1.23}$ ($-\num{68}$\% std.) \\ 
& OneMfrOnly (3/20) & $+\num{0.57}$  & $+\num{4.82}$ & $+\num{2.01}$ ($-\num{34}$\% std.) \\ 
\midrule
\multirow{5}{*}{\rotatebox{90}{\textbf{HuBERT}}}
& OnePerMfr (10/20) & $-\num{0.43}$ & $+\num{6.83}$   & $+\num{1.33}$ ($-\num{61}$\% std.) \\ 
& FirstGen (5/20) & $-\num{0.17}$ & \underline{$+\num{9.12}$} & $+\num{1.55}$ ($-\num{80}$\% std.)\\
& Pre2022 (17/20) & $-\num{0.41}$ & \underline{$+\num{7.92}$}  & $+\num{0.93}$ ($-\num{73}$\% std.) \\ 
& AllButOne (19/20) & $-\num{0.41}$  & \underline{$+\num{8.43}$} & $+\num{1.15}$ ($-\num{73}$\% std.)\\ 
& OneMfrOnly (3/20) & $-\num{0.24}$ & $+\num{4.62}$  & $+\num{1.29}$ ($-\num{40}$\% std.) \\ 
\hline
\end{tabular}}
\end{table}
Regardless of the device subset used for training or the speech encoder type, all BT 
models improve the UAR over their respective baselines when averaged across all microphone conditions (Table~\ref{performance_tab} condition \textit{All}).
Furthermore, the BT models maintain comparable performance on the untransformed condition, with only slight degradations 
(or even gains for some \mbox{Wav2Vec-based} models) as shown in Table \ref{performance_tab} (\textit{Base} condition).
The absolute gain of our approach tends to increase with the severity of the original performance degradation under the respective microphone condition.
For example, under the worst-case microphone condition (the one with the largest baseline performance gap), improvements reach up to \qty{9.12}{pp} (Table~\ref{performance_tab} condition \textit{Worst}).  
Notably, the BT models developed using the OnePerMfr subset achieve improvements of \num{7.07}/\qty{6.83}{pp} (condition \textit{Worst}), despite the corresponding device (Moto G9) being unseen during training, indicating strong generalisation potential to unseen but challenging microphones.  
Considering the overall gain (condition \textit{All}), training subsets excluding the Moto G9 device (OnePerMfr and OneMfrOnly), 
yield the largest improvements and also achieve the best performance on the untransformed condition for the \mbox{Wav2Vec-based} models.
In contrast, when the Moto G9 device is included in the training set (subsets FirstGen, Pre2022, AllButOne), enforcing invariance to this challenging device appears to slightly
harm the generalisation to others. 
For \mbox{HuBERT-based} models, utilizing the FirstGen subset yields the highest overall gain.
For this particular model, encouraging invariance to the challenging device condition does not appear to come at the expense of the reduced overall gain observed for the \mbox{Wav2Vec-based}
models and the other \mbox{HuBERT-based} models.
In all cases, the BT module stabilises the performance across 
conditions compared to the baselines, as it reduces the standard deviation by up to 80\% (Table~\ref{performance_tab}).
\begin{figure}[]
  \includegraphics[width=\columnwidth]{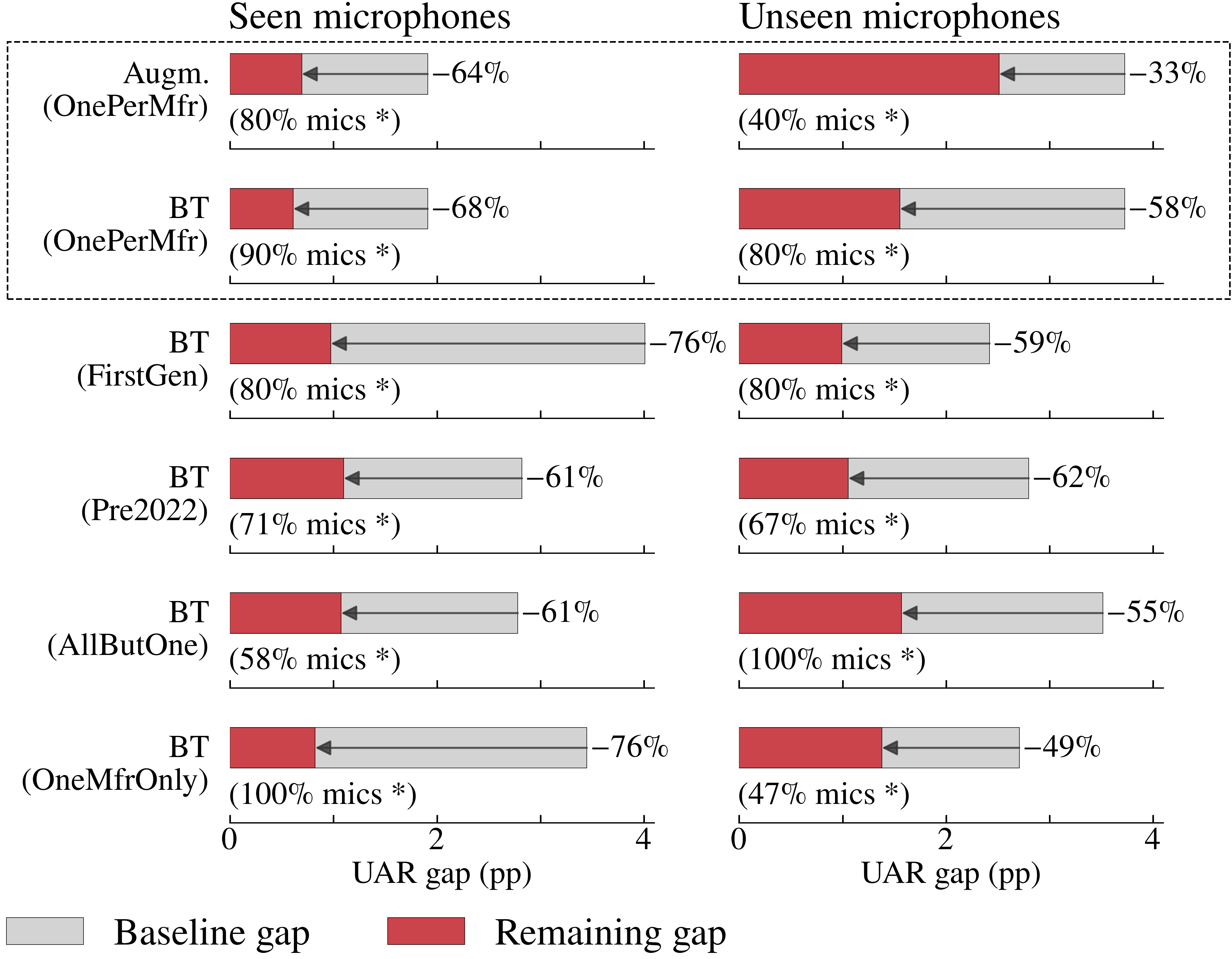}
  \caption[width=\columnwidth]{Performance gaps in UAR for the Wav2Vec baseline (gray) and the remaining gap for the respective Barlow Twins (BT) model (red). Results are shown separately for seen (left) and unseen (right) microphones with respect to the BT training. The training device subset is noted in parentheses for the BT models (see Section~\ref{sec:device_subsets} for details on device splits).
  Arrows show the absolute gap reduction, while the percentage beside each bar indicates the relative gap reduction. The proportion of microphones with significant gap reductions is indicated as \mbox{X\% mics *}. For comparison, gap reductions obtained with the \mbox{augmentation baseline (Augm.)} trained with subset OnePerMfr are shown, directly comparable to the results of the BT model (OnePerMfr), as indicated by the dashed line.}
  \label{robustness_analysis_fig}
\end{figure}
For additional insight, 
Figure~\ref{robustness_analysis_fig} focuses on absolute and relative gap reductions between the \mbox{Wav2Vec-based} baseline and the respective BT models, while aggregating results over seen and unseen microphone conditions separately.
By using our approach, the baseline gap is reduced by up to \num{76}\% across seen conditions.
The improvement on unseen microphones is smaller (up to \num{62}\% gap reduction) but quite consistent across device subsets and significant for most unseen microphones in four of the five splits.
This indicates generalisation capabilities to diverse types of microphone mismatch as motivated in Section~\ref{sec:device_subsets}. 
Comparing the different BT models, those trained on subsets with less than \num{10} devices (FirstGen, OneMfrOnly) achieve the highest gap reductions across seen conditions, 
potentially reflecting the more frequent exposure to the characteristics of each individual device during training.
However, they also exhibit the largest seen-unseen difference in gap reduction. 
In contrast, models trained on subsets with larger numbers of devices (OnePerMfr, Pre2022, AllButOne) show a more balanced benefit across seen and unseen devices.
Compared to basic data augmentation (Augm. model, Figure~\ref{robustness_analysis_fig}), 
the (\mbox{Wav2Vec-based}) BT model (OnePerMfr) achieves similar performance on seen microphones but nearly double the improvement on unseen ones.
This demonstrates the benefit of encouraging explicit microphone invariance in the representation space.
We additionally evaluated our approach against an adversarial baseline following Wang et al.~\cite{wang2026domain}. The BT models consistently outperformed it on the unseen condition,
achieving, on average, a 22\% greater gap reduction than the adversarial model
(further results omitted due to space constraints).

\subsection{Embedding space analysis}

\begin{figure}[]
  \includegraphics[width=\columnwidth]{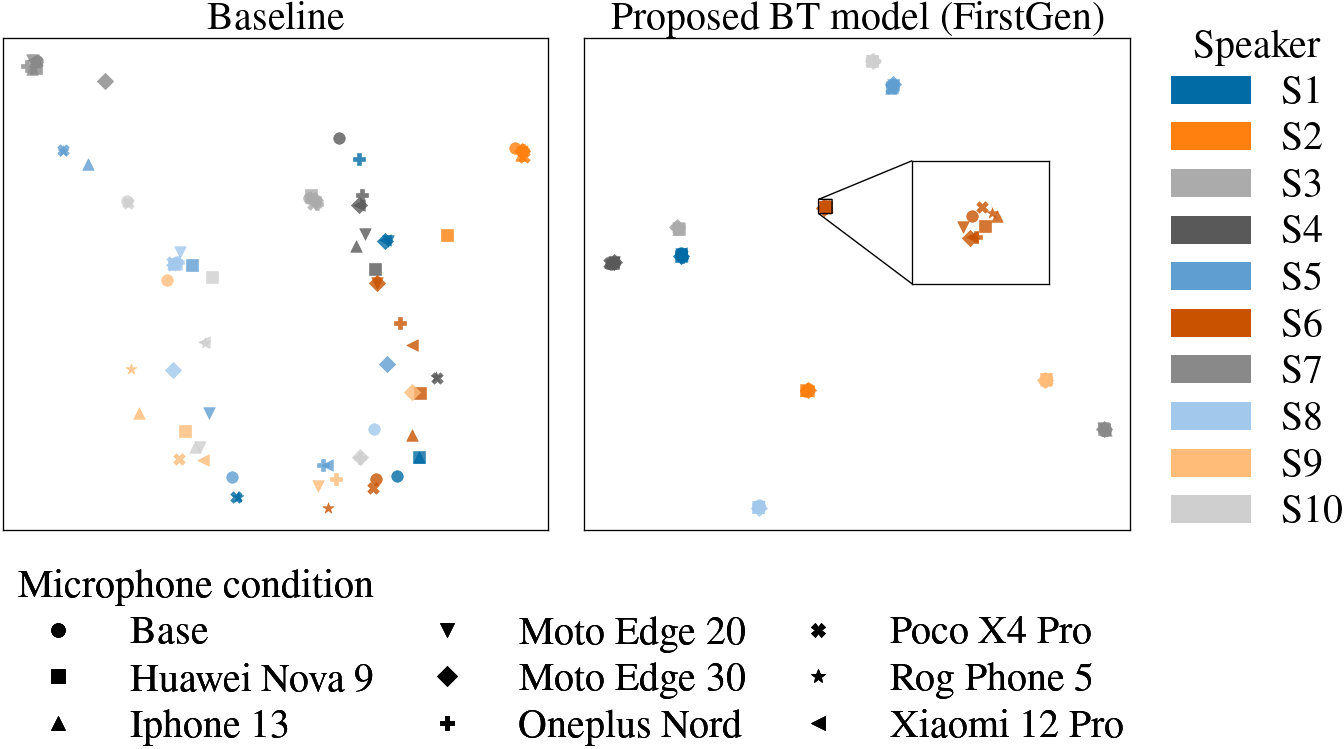}
  \caption[width=\columnwidth]{Visualisation of 2D \mbox{t-SNE} embeddings from 10 speakers.
  Baseline embeddings and those from the proposed \mbox{Wav2Vec-based} Barlow Twins (BT) model (trained with subset FirstGen) are shown. Only microphone conditions unseen during BT model training and the untransformed (\textit{Base}) condition, are included.}
  \label{embedding_space_fig}
\end{figure}

Besides the impact on downstream performance, we also examine the direct effect of the BT module on the embedding space \mbox{(post-pooling)}.
As shown in Figure~\ref{embedding_space_fig}, baseline embeddings of the same speech utterance from a speaker under different microphone conditions, are widely spread, with heavy speaker overlap.
In contrast, the BT models (shown for \mbox{Wav2Vec-based} BT model (FirstGen)) form compact, \mbox{non-overlapping} speaker clusters.
This effect holds for HuBERT-based models, suggesting successful microphone invariance enforcement at the embedding-level.

\section{Conclusion}
\label{sec:conclusion}

To the best of our knowledge, this study presents the first analysis of microphone mismatch in \mbox{transformer-based} voice disorder detection, focusing on smartphones.
This is particularly important, as smartphones represent a widely used platform for digital public health tools and \mbox{long-term} health monitoring applications.
While most microphones cause moderate degradation, individual
ones can have a substantial impact. 
To address this, we integrate a \mbox{BT-inspired} module to enforce \mbox{microphone-invariant} embeddings. 
Our method improves the structure of the embedding space and significantly reduces the performance gap for most considered microphones. 
The analysis suggests generalisation to unseen devices across manufacturers, smartphone generations and release periods. 
Training with different device subsets reveals a \mbox{trade-off} between robustness to specific challenging devices and broader
generalisation, underscoring the importance of the subset design.
However, due to the lack of \mbox{multi-microphone} pathological
data, results are based on simulated microphone characteristics and require further validation.

\newpage
\section{Acknowledgements}

This work has been in part funded by the Deutsche Forschungsgemeinschaft (DFG, German Research Foundation) - 459360854 as part of the Research Unit "Lifespan AI: From Longitudinal Data to Lifespan Inference in Health" (DFG FOR 5347) and
the Leibniz ScienceCampus Bremen Digital Public Health, which is jointly funded by the Leibniz Association (W72/2022), the Federal State of Bremen, and the Leibniz Institute for Prevention Research and Epidemiology – BIPS.

\label{sec:refs}

\bibliographystyle{IEEEbib}
\bibliography{refs}

\end{document}